\documentclass[aps,prl,showpacs,twocolumn,groupedaddress,amsmath,amssymb]{revtex4-1}
\usepackage{times,xspace}
\usepackage{amsfonts}
\usepackage{amsfonts}
\usepackage{bm}
\usepackage{amssymb}
\usepackage[final]{graphicx}
\usepackage{color}
\begin{document}

\title{Modulated superconductivity due to vacancy and magnetic order in $A_x$Fe$_{2-x/2}$Se$_2$ [$A=$Cs, K, (Tl,Rb), (Tl,K)] iron-selenide superconductors}

\author{Tanmoy Das$^1$, and A. V. Balatsky$^{1,2}$}
\affiliation{$^1$Theoretical Division, Los Alamos National Laboratory, Los Alamos, NM, 87545, USA.\\
$^2$Center for Integrated Nanotechnologies, Los Alamos National Laboratory, Los Alamos, NM, 87545, USA.}

\date{\today}
\begin{abstract}
We present a calculation of a `modulated' superconducting state in iron-selenide superconductors. The zero-momentum $d-$wave pairing breaks the translational symmetry of the conventional BaFe$_2$Se$_2$-like crystal of $I4/mmm$ space group. This pairing state becomes compatible when the Fe vacancies form an ordered state and the crystal symmetry changes to a low-temperature $I4/m$ one. For the specific case of an incommensurate vacancy order at $Q_v=(\frac{1}{5},\frac{3}{5})$ in K$_{0.82(2)}$Fe$_{1.626(3)}$Se$_2$, we find that it induces a block checkerboard antiferromagnetic phase at wavevector ${\bf Q}_m=4{\bf Q}_v$. The coexistence of vacancy order and magnetic order leads to a reconstructed ground state which naturally couples to the $d-$wave superconductivity in a uniform phase in what we propose will be a general coupling for all iron-selenide superconductors.  Our results agree with numerous experimental data available to date. We thus suggest that the incommensurability leads to uniform coexistence of multiple phases as a viable alternative to a nanoscale phase separation  in high-$T_c$ superconductors and play an important role in the enhancement of superconductivity.
\end{abstract}
 \maketitle\narrowtext
{\it Introduction$-$}The layered iron-selenide compounds $A_x$Fe$_{2-x/2}$Se$_2$ [$A=$Cs, K, (Tl,Rb), (Tl,K)] have been found recently to become superconducting (SC) with highest $T_c\sim32$K when a suitable value of the Fe vacancy concentration $x\sim0.87$ is achieved.\cite{WBao_phase}
%Upon adding Fe into the vacancies, the insulating $A_x$Fe$_{2-x/2}$Se$_2$ [$A$=K or Cs] can be made superconducting at a critical %value of the vacancy level $x\sim0.87$.
This fascinating tunability of the SC  property with Fe vacancy has opened up a new platform to study the mechanism of unconventional superconductivity.  Most interestingly, the onset of vacancy order is accompanied by several interesting changes of the structural, electronic, magnetic and SC properties. In particular, numerous experiments have found that the superconductivity turns on exactly when the randomly placed Fe vacancies also form an ordered state in addition to a structural phase transition.\cite{WBao_Tv} Simultaneously, the value of the magnetic moment is dramatically enhanced on each Fe sites.\cite{WBao_Tv}

We show here that in these materials the vacancy order induces the magnetic order, and that both phases uniformly coexist with $d$-wave superconductivity to lead to a `modulated superconductor'. The induction of a modulated order due to the incommensurability of another, competing,  order is a common phenomenon in unconventional superconductors. Historically the incommensurate SDW order in cuprates was first found in neutron scattering experiment La$_{1.6-x}$Nd$_{0.4}$Sr$_x$CuO$_4$ with wavevector $Q_m=(\pi\pm\delta,\pi)/(\pi,\pi\pm\delta)$,\cite{tranquada} similar to the case of Cr.\cite{Fawcett} Subsequently, an induced CDW order at $Q_c=(2\delta,0)/(0,2\delta)$ is also measured in scanning tunneling microscopy (STM).\cite{EHudson,Dastwogapcuprate} Coexistence of SDW and CDW produces a `stripe'-like pattern in cuprates which is observed by neutron scattering and ARPES in La-based cuprate, STM in Bi2201 and quantum oscillations, Nernst  signal and Seeback measurements in YBCO families.\cite{Vojta,Bob,DasNFL} Similarly, in iron-pnictides (BaFe$_2$As$_2$) and chalcogenides (FeTe$_{1-x}$Se$_x$), the SDW with a modulation $Q_m=(\pi,0)/(0,\pi)$ has been argued  to lead to a CDW with $Q_c=2Q_s$.\cite{Sasha_pnictide} The phenomenology also extends to the `hidden-order' state of heavy fermion URu$_2$Si$_2$ which has an incommensurate modulation at $Q_{h}=(0.6\pi,0)/(0,0.6\pi)$ and the formation of the CDW at $Q_c=2Q_h$ has been predicted.\cite{Sasha_URuSi} It is thus timely and important to investigate the nature of the density wave order of the conventional SC  quasiparticles, and their coupling to the vacancy order in Fe-selenide as another example of same basic principle of modulated states high-$T_c$ superconductivity.

\begin{figure*}%[h]
\hspace{-0cm}
%\rotatebox{0}{\scalebox{.47}{\includegraphics{SC_gap_symmetry_cartoon_v5_Page_1.jpg}}}
\rotatebox[origin=c]{0}{\includegraphics[width=1.9\columnwidth]{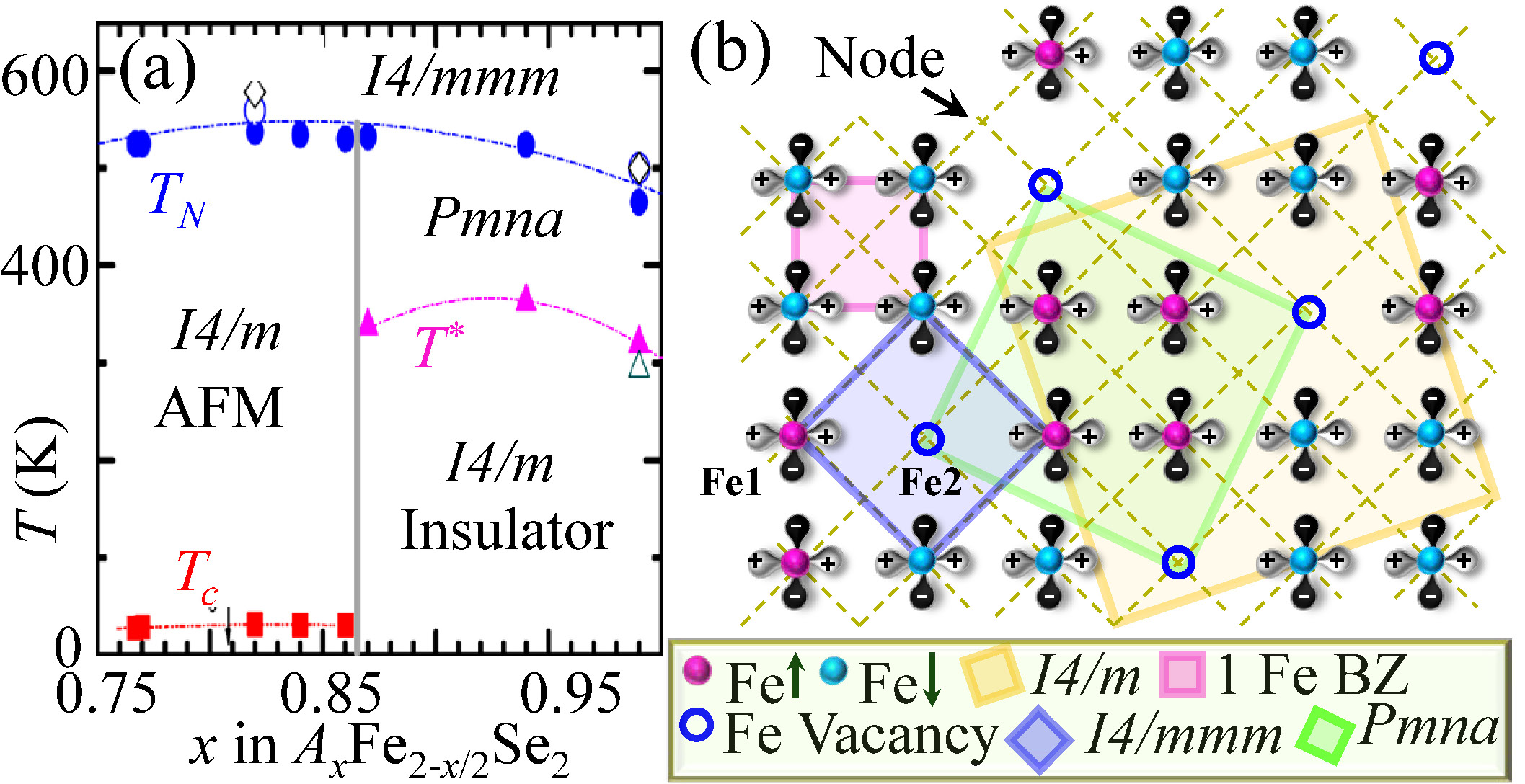}}
\caption{(Color online) (a) Experimental phase diagram of A$_x$Fe$_{2-x/2}$Se$_2$ compound taken from Ref.~\cite{WBao_phase} is also shown to be a common phase diagram for this class of materials $A_x$Fe$_{2-x/2}$Se$_2$ in Ref.~\cite{WBao_common}. At high temperature, the system is in the Ba$_2$Fe$_2$As$_2$-like tetragonal structure of $I4/mmm$ space group at all Fe vacancies. A tetragonal structure with $I4/m$ unit cell is observed to stabilize at low temperature in the $\sqrt{5}\times\sqrt{5}\times1$ superlattice. The metal-insulator transition is observed around $x\sim0.865$ in numerous experiments albeit a smooth behavior of the N\'eel temperature $T_N$ (diamond) throughout the phase diagram. The open diamond symbol marks the transition temperature at which disorder to order state of the Fe vacancy is measured. Solid squares represent $T_c\sim30$K which seem to be symmetric around $x=0.8$ at which full occupancy of Fe2 sublattices and vacant Fe1 sites is realized. (b) Common crystal and magnetic structure of $A_2$Fe$_4$Se$_5$ in the real space tetragonal $I4/m$ unit cell (top view), only Fe atoms are shown.\cite{WBao_Tv,WBao_phase,WBao_common} Magenta and cyan color circles give spin-up and spin-down configurations, while black and black dumbbells represent positive and negative phase of the pairing symmetry. Boxes of different colors and sizes stand for different unit cells (text), while yellow box is the true unit cell for this system at low-temperature.}\label{fig1}
\end{figure*}
 Fig.~1(a) gives the experimental phase diagram\cite{WBao_phase} as a function of temperature and Fe vacancy for $A_x$Fe$_{2-x/2}$Se$_2$.  Our calculations consistently demonstrate that the vacancy order in Fe-selenide at a superlattice modulation of ${\bf Q}_v=(\frac{1}{5},\frac{3}{5})$\cite{TEM,WBao_common} breaks the translational symmetry of the tetragonal crystal in such a way that it induces structural, magnetic and unconventional SC phase transition exactly at the same value of $x\sim0.87$. At the vacancy order transition temperature $T_v\simeq$578K in K$_{0.82(2)}$Fe$_{1.626(3)}$Se$_2$\cite{WBao_Tv}, the crystal structure undergoes a transition from the high-temperature tetragonal BaFe$_2$As$_2$-like structure of $I4/mmm$ space group to a lower symmetry tetragonal structure of $I4/m$ where the latter can be understood as a $\sqrt{5}\times\sqrt{5}\times1$ superlattice of the former, see Fig.~1(b) Subsequently, a block checkerboard like antiferromagnetic (AFM) order stabilized with significantly enhanced moments on the Fe sites [about 3.3$\mu_B$] below a N\'eel temperature $T_N\approx559$K$<T_v$ with an order vector ${\bf Q}_m=(\frac{4}{5},\frac{2}{5})$.\cite{WBao_Tv} Finally, the superconductivity appears below $T_c\simeq32$K. Later on this overall phenomena has been observed to be a common feature in all iron-selenide chalcogenides studied so far.\cite{WBao_common,WBao_phase}

{\it Vacancy order and magnetic order$-$} We concentrate here at a particular case of K$_{0.82(2)}$Fe$_{1.626(3)}$Se$_2$\cite{WBao_Tv} at which the vacancy superstructure and the magnetic structure have propagation vectors ${\bf Q}_v=(\frac{1}{5},\frac{3}{5})$ and ${\bf Q}_m=(\frac{4}{5},\frac{2}{5})$. The property of the magnetic order $Q_m=4Q_v$ motivates us to study a modulated order model in which the vacancy order
%with $T_v=578$ K
 is the driving instability which induces the magnetic order at $T_N<T_v$. Vacancy order will naturally develop a CDW at $2Q_v$. Neutron and x-ray study in a Cs intercalated compound Cs$_y$Ke$_{2-x/2}$Se$_2$ showed the presence of the superstructure reflection at a $Q=(\frac{2}{5},\frac{1}{5})$\cite{Pomjakushin2Q} which is equal to $2Q_v$ (antiphase).  Finally, the vacancy order clearly breaks the rotational $C_4$ symmetry of the crystal, % down to $C_2$,
 leading to a global nematic order parameter. The unidirectional nematic order breaks the $t_{2g}$ symmetry of the Fe $d$ orbitals (which contribute mostly to the FS) and thus an orbital order around each vacancy can also be expected. We assume the orbital order will have modulation at $3Q_v$. Such an induction process is reversible as long as the modulation vector is incommensurate which we assume to be the case. It is interesting to note that such possible nematic order is consistent with claims of nematicity in cuprates,\cite{Vojta}, pnictide\cite{Sasha_pnictide}, URu$_2$Si$_2$\cite{Varma_URuSi,Sasha_URuSi} and  Sr$_3$Ru$_2$O$_7$\cite{Borzi_SrRuO}, although the details of the ordering are material sensitive.

The coupling between the modulated order with $d-$wave superconductivity are strongly constrained by 
symmetry and momentum conservations and belong to a point-group-symmetry. Due to $C_4$ symmetry breaking, $s$-wave and $d$-wave representations always mix, but $d_{x^2-y^2}$ and $d_{xy}$ pairing remain clearly distinguishable.\cite{Vojta,Sachdev} Therefore, the order parameter can be defined globally via the electronic momentum distributions as $\phi_g=\sum_{{\bf k}}d_{\bf k}\langle c^{\dag}_{{\bf k}^{\prime},\alpha}{\bf \Sigma}_{\alpha,\beta}c_{{\bf k}\beta}\rangle$, where $d_{\bf k}$ is the structure factor for various $d-$waves and ${\Sigma}_{\alpha,\beta}=\delta_{\alpha,\beta}$ or $\sigma_{\alpha,\beta}$ for scaler and vector order, respectively. Decoupling the global parameter into the local orders, one can acquire well defined directions as $\phi_i=\sum_{{\bf k},\sigma}\langle c^{\dag}_{{\bf k}+{\bf Q}_i\sigma}c_{{\bf k}\sigma}\rangle$ for scalar or $\phi_i=\sum_{{\bf k},\sigma}\langle c^{\dag}_{{\bf k}+{\bf Q}_i\uparrow}c_{{\bf k}\downarrow}\rangle$ for spin orders. Similarly for superconductivity, we obtain modulated orders $\psi_i=\sum_{\bf k}\langle c^{\dag}_{{\bf k}+{\bf Q}_i\uparrow}c^{\dag}_{-{\bf k}\downarrow}\rangle$, while the zero-momentum SC order parameter is $\psi_0=\sum_{\bf k}\langle c^{\dag}_{{\bf k}\uparrow}c^{\dag}_{-{\bf k}\downarrow}\rangle$. 
%In general, the directional dependence of order parameters can be decomposed according to the representation of the point group.
In the present case, we restrict ourselves to case with $i=1-4$:  $i=1$ for vacancy order, $i=2$ for charge order, $i=3$ for orbital order and $i=4$ for magnetic order. Keeping only the linear terms in the global modulated order parameter $\phi_n$ in the tetragonal crystal environment, we can write all the coupling terms\cite{Vojta}
\begin{eqnarray}\label{eq:coupling}
&&\phi_g\sum_{i=1}^4\lambda_i|\phi_{i\alpha}|^2 + \phi_1^*\sum_{i=2}^4\lambda_{1i}\phi_1^*\phi_i^2\nonumber\\
&&+\sum_{i=1}^n\left[\lambda^s_0\phi_1^*\psi_i{\bar{\psi_i^*}}+\lambda^s_i(\phi_1^*\psi_0{{\psi_i^*}}+\phi_1\psi_0{\bar{\psi_i^*}})  ... + c.c \right].
\end{eqnarray}
%
% and $\psi_0$, $\psi_i (\bar{\psi_i})$ are the uniform and modulated SC condensate order parameters. 
The coupling constants $\lambda_{1i}$ between vacancy order with others are possible only if the corresponding modulation follows ${\bf Q}_i=n{\bf Q}_1$ relation (${\bf Q}_1={\bf Q}_v$). We will comment on the superconductivity below.

In Fig.~2, we show the evolution of the FS reconstruction and the nature of the gap opening in the normal state. Absorbing the coupling constant in the order parameters, we define the mean-field gaps as $V_i~(=\lambda_i\phi_i)$. The eigenvector is $\Psi^{\dag}_0=\big(c^{\dag}_{{\bf k},\sigma}~c^{\dag}_{{\bf k}+{\bf Q}_1,\sigma}~ c^{\dag}_{{\bf k}+{\bf Q}_2,\sigma}, ...\big)$, where $c^{\dag}_{{\bf k}+{\bf Q}_1,\sigma}$ is the creation operator for the $Q_1$th order and so on. We construct the two band model Hamiltonian in 2 Fe unit cell in the normal state [i.e. when SC order parameters $\psi_0$ and $\psi$ vanish] as

%%%%%%%%%%%%%%%%
\begin{widetext}
\begin{eqnarray}
H_0=\Psi^{\dag}_{0}\left(
\begin{array}{ccccccccccccccccccccc}
% line 1
\xi^{11}_{\bf k} & \xi^{12}_{\bf k} & V_{v} & 0 & V_{2}& 0 & V_{3} & 0 & V_{m} & 0
&\ldots & V_{m} & 0 & V_{3} & 0 & V_{2} & 0 & V_{v}& 0\\
% line 2
\xi^{21}_{\bf k} & \xi^{22}_{\bf k} & 0 & V_{v} & 0 & V_{2} & 0 & V_{3} & 0 & V_{m}
& \ldots & 0 & V_{m}& 0 & V_{3} & 0 & V_{2} & 0 & V_{v}\\
% line 3
V_{v}^* & 0 & \xi^{11}_{{\bf k}+{\bf Q}} & \xi^{12}_{{\bf k}+{\bf Q}} & V_{v} & 0 & V_{2} & 0 & V_{3} & 0
&\ldots & 0 &0& V_{m} & 0 & V_{3} & 0 & V_{2} & 0\\
%line 4
0 & V_{v}^* & \xi^{21}_{{\bf k}+{\bf Q}} & \xi^{22}_{{\bf k}+{\bf Q}} & 0 & V_{v} & 0 & V_{2} & 0 & V_{3}
& \ldots & 0 &0 &0 & V_{m} & 0 & V_{3} & 0 & V_{2}\\
%line 5
\vdots & \vdots & \vdots & \vdots &\vdots & \vdots & \vdots & \vdots &\vdots & \vdots &\ddots
& \vdots & \vdots & \vdots & \vdots &\vdots & \vdots & \vdots & \vdots \\
V_{v}^* & 0 & V_{2}^* & 0 & V_{3}^* & 0 & V_{m}^* & 0 & 0 & 0 & \ldots&V_{3}^* & 0&V_{2}^* & 0&V_{v}^* & 0 & \xi^{11}_{{\bf k}+9{\bf Q}} & \xi^{12}_{{\bf k}+9{\bf Q}}\\
0&V_{v}^* &0&V_{2}^* & 0 & V_{3}^* & 0 & V_{m}^* & 0 & 0 & \ldots&0 & V_{3}^*&0&V_{2}^*&0&V_{v}^*&\xi^{21}_{{\bf k}+9{\bf Q}} & \xi^{22}_{{\bf k}+9{\bf Q}}
\end{array}
\right)\Psi_{0}.
\end{eqnarray}
\end{widetext}
%%%%%%%%%%%%
%
Bare dispersion, $\xi_k$, matrix arises from two tight-binding bands of strongly hybridized $t_{2g}-$orbitals present near the Fermi-level.\cite{Das_FeSe} In such incommensurate case, the FS reconstruction is determined by  multiple translations of the bare band structure $\xi_{\bf k}$ by $\pm n{\bf Q}_v$, where the largest value of the integer $n$ is determined by ${\bf Q}_v=(2\pi,2\pi)$ in 2D which is 10 in our case. Each component of the translated dispersion is denoted by $\xi_{{\bf k}+n{\bf Q}}$. As in the case of non-SC Cr,\cite{Fawcett} cuprate\cite{Sachdev} or other SC materials\cite{Sasha_pnictide,Sasha_URuSi,Varma_URuSi,Borzi_SrRuO} incommensurability causes hierarchy of gap openings of order $\Delta^{\nu,\nu^{\prime}}_m\sim(V_v^{\nu m}/W^{m-1}+V_v^{\nu^{\prime} m}/W^{m-1})$ [$W$ is bare band width, $\Delta$ should not be confused with SC gap] at band crossings between $\xi^{\nu}_{{\bf k}+n{\bf Q}}$ and $\xi^{\nu^{\prime}}_{{\bf k}+(n\pm m){\bf Q}}$.\cite{foot3}  As long as $V_{i}<<W$, the FS is well described by including the lowest-order gap only and we will neglect all the matrix elements with $m>1$ and the momentum dependence in the potentials $V_i$.

\begin{figure*}%[h]
\hspace{-0cm}
%\rotatebox{0}{\scalebox{.45}{\includegraphics{SC_gap_symmetry_cartoon_v5_Page_2.jpg}}}
\rotatebox[origin=c]{0}{\includegraphics[width=1.95\columnwidth]{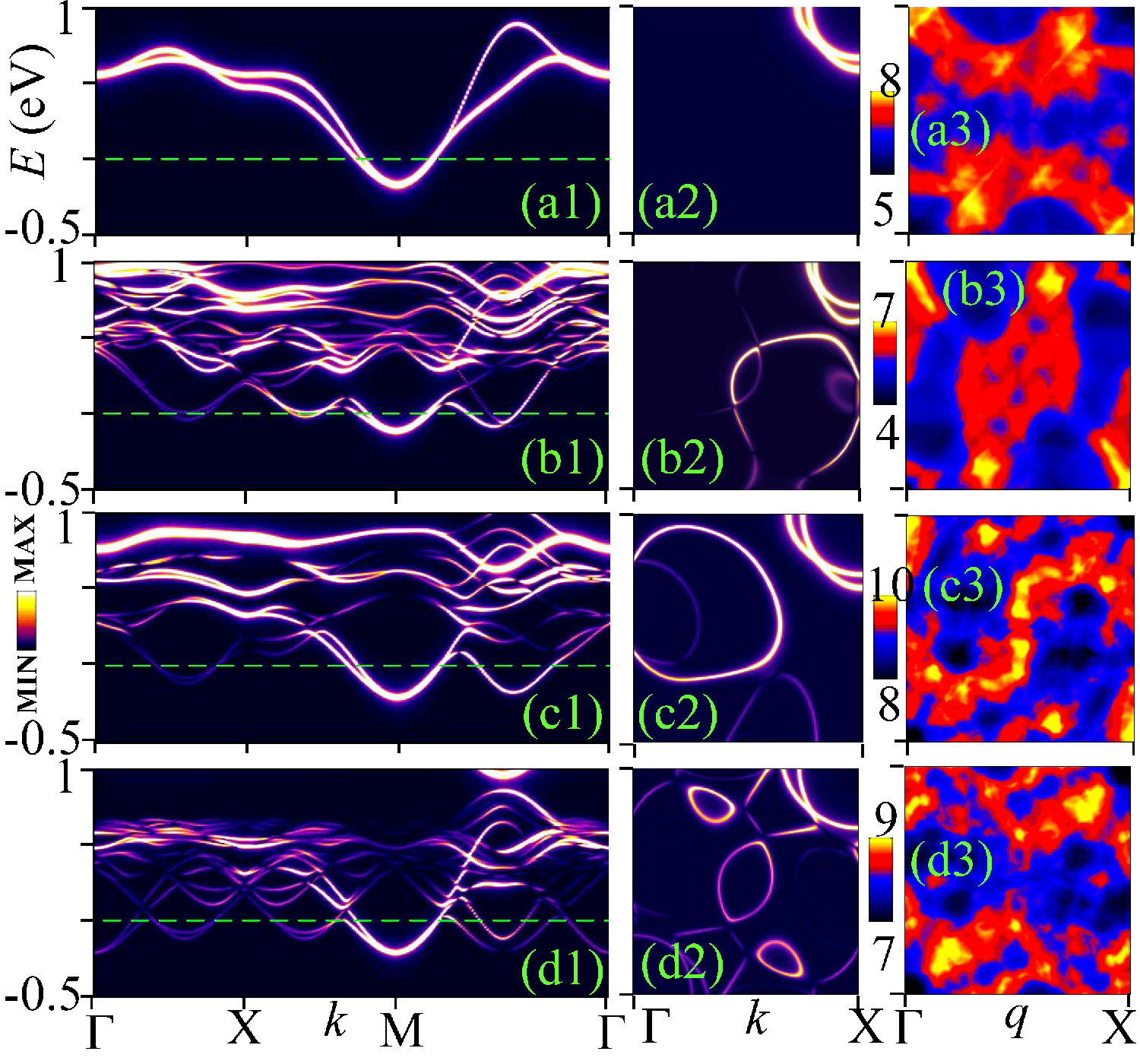}}
\caption{(Color online) The band dispersion along high-symmetry lines [(a1),(b1),(c1),(d1)], corresponding FSs [(a2),(b2),(c2),(d2)] and bare susceptibility [(a3),(b3),(c3),(d3)] maps at zero energy are presented for four combinations of interactions: (a)$\rightarrow$ all $V_i=0$,  (b)$\rightarrow$ only $V_v=-0.5$eV, (c)$\rightarrow$ only $V_m=0.02$eV and (d)$\rightarrow$ $V_v=-0.5, V_2=0.2, V_3=-0.1, V_m=0.01$eV. FS and susceptibility are plotted only in the first quadrant of the BZ which are symmetric in the other quadrants. Note that opposite phases between consecutive interactions give maximal weight on main bands.  Note that the electronic structure consists of 20 split bands while the spectral weight is largest in the main band if coherence factors of the modulated order are included. Here we have not included the structure factor of the modulated order which may further enhance the distribution of the spectral weight among the modulated bands.
}\label{fig1}
\end{figure*}

We have used the eigenstates and eigenvectors of Eq.~2 to compute the band dispersion along high-symmetry lines and the corresponding FS in the normal state, as shown in Fig.~2. The results are presented for various $V_i$, while fixing the number of quasiparticles on the FS. We also present the real part of the non-interacting static susceptibility in the 2D $q$-space  ignoring matrix-element effects.

To understand the evolution of the modulated order due to turning on of individual interactions, it is useful to examine them first separately. In the nonordered state (all $V_i=0$), the FS consists of two electron-like closed orbits centering $M-$point in the 2 Fe unit cell, see Figs.~2(a1,a2). The corresponding susceptibility in Fig.~2(a3) shows nesting peaks or streaks at all $nQ_v$s while the intensity dominates at $Q_v$ and $Q_m=4Q_v$. With turning on of $V_v=-0.5$eV, the vacancy order is illuminated and the FS reconstruction becomes complicated, see Figs.~3{b1,b2,b3}. While both electron pockets are still present in closed orbits, but their size and the associated spectral weight are reduced. The missing spectral weight is shared by a newly developed electron-pocket and several open orbit FS pieces. The phase of the interaction $V_v$ does not have significant effect on modulated pattern (not shown).

We now study the magnetic ordering in Fig.~2(c). To connect the magnetic order parameter $V_m$ with experiments we will investigate local moments. Neutron diffraction has determined a large magnetic moment block checkerboard AFM in the range of 2.3-3.3$\mu_B$/Fe for all known iron-selenide superconductors.\cite{WBao_Tv,WBao_common} Within our mean-field approach, we find that a value of $V_m=20$meV produces maximal local moments of $2\mu_B\langle S_z\rangle_{max}\approx0.12\mu_B$ in 2 Fe unit cell which is smaller than seen in experiments. The value of computed magnetic moments can be enhanced by various modifications of our calculation, e.g. when all $d-$ orbitals are included.    With all interactions included, a drastic band folding and FS reconstruction are observed in Figs.~2(d). The corresponding susceptibility continues to exhibit dominant nesting along the vacancy order vector.  First principle calculation have  demonstrated drastic FS reconstruction driven by the Fe vacancy and AFM order.\cite{caoFSrecon} Such a FS reconstruction is consistent with observations such as transmission electron microscopy (TEM),\cite{TEM} NMR\cite{WYu_NMR}, Neutron diffraction\cite{WBao_Tv} and optical studies.\cite{optical}

{\it Vacancy order and $d$-wave pairing.$-$} As superconductivity turns on, the coupling between the competing order to SC $\lambda^s_i$ in Eq.~1 are allowed only when ${\bf Q}_1=i{\bf Q}_s$ where $i=1,2 ...$ (${\bf Q}_s$ is the SC ordering vector in the pair-density channel). Interestingly, the zero-momentum condensate ${\bf Q}_s=0$ is always a complimentary component of the SC state, whereas the finite momentum components are subject to the higher order coupling strength and depend on the nature of the driving instability vector ${\bf Q}_1$. Note that $\lambda^s_1$ alone generates a Fulde-Ferrell-Larkin-Ovchinnikov state.\cite{footnematic} Absorbing the coupling constants $\lambda^s_i$ in the order parameter, we define the SC gaps as $\Delta_i=\lambda_i^s\psi_i$.  Similar to Eq.~2, we can construct the SC gap matrix $H_{\rm SC} = \big(\Delta_{\bf k}~ \Delta_{{\bf k}+{\bf Q}_1} ~\Delta_{{\bf k}+{\bf Q}_2}~...; c.c.\big)$. In the Nambu space defined by $\Psi^{\dag}({\bf k})=\big(\Psi^{\dag}_0({\bf k}); \Psi^{\dag}_0(-{\bf k})\big)$, we get the modulated SC Hamiltonian as
\begin{eqnarray}
H=\Psi^{\dag}\left(
\begin{array}{cc}
% line 1
H_0({\bf k}) & H_{\rm SC}({\bf k})\\
% line 2
H^*_{\rm SC}(-{\bf k}) & H_0(-{\bf k})\\
\end{array}
\right)\Psi.
\end{eqnarray}
%

%$H = \big(H_0({\bf k}) ~H_{\rm SC}({\bf k}); H^*_{\rm SC}(-{\bf k}) ~H_0(-{\bf k})\big)$.

In the rest of the paper we discussion how the vacancy order play an important role to the onset of $d-$wave superconductivity. It has been argued that the $d-$wave order in these materials without a hole-pocket at $\Gamma$ in the conventional BaFe$_2$As$_2$-like crystal structure breaks the translational symmetry.\cite{Das_FeSe,Mazin} To elucidate this, we recall first the example of BaFe$_2$As$_2$ pnictide superconductor which hosts both hole and electron pockets at $\Gamma$ and M points, respectively, in the 2 Fe unit cell and how $s^{\pm}$-pairing is compatible with the corresponding crystal symmetry. We illustrate this with the help of the unitary transformation from 1 Fe unit cell to the 2 Fe unit cell where one expects, by construction, to retain all observables such as nodeless SC gap(s) and FS nesting  to be compatible with the actual crystal symmetry irrespective of the conventional one used for calculations.
\begin{figure*}%[top]
\hspace{-0cm}
%\rotatebox{0}{\scalebox{.37}{\includegraphics{SC_gap_symmetry_cartoon_v5_Page_3.jpg}}}
\rotatebox[origin=c]{0}{\includegraphics[width=1.95\columnwidth]{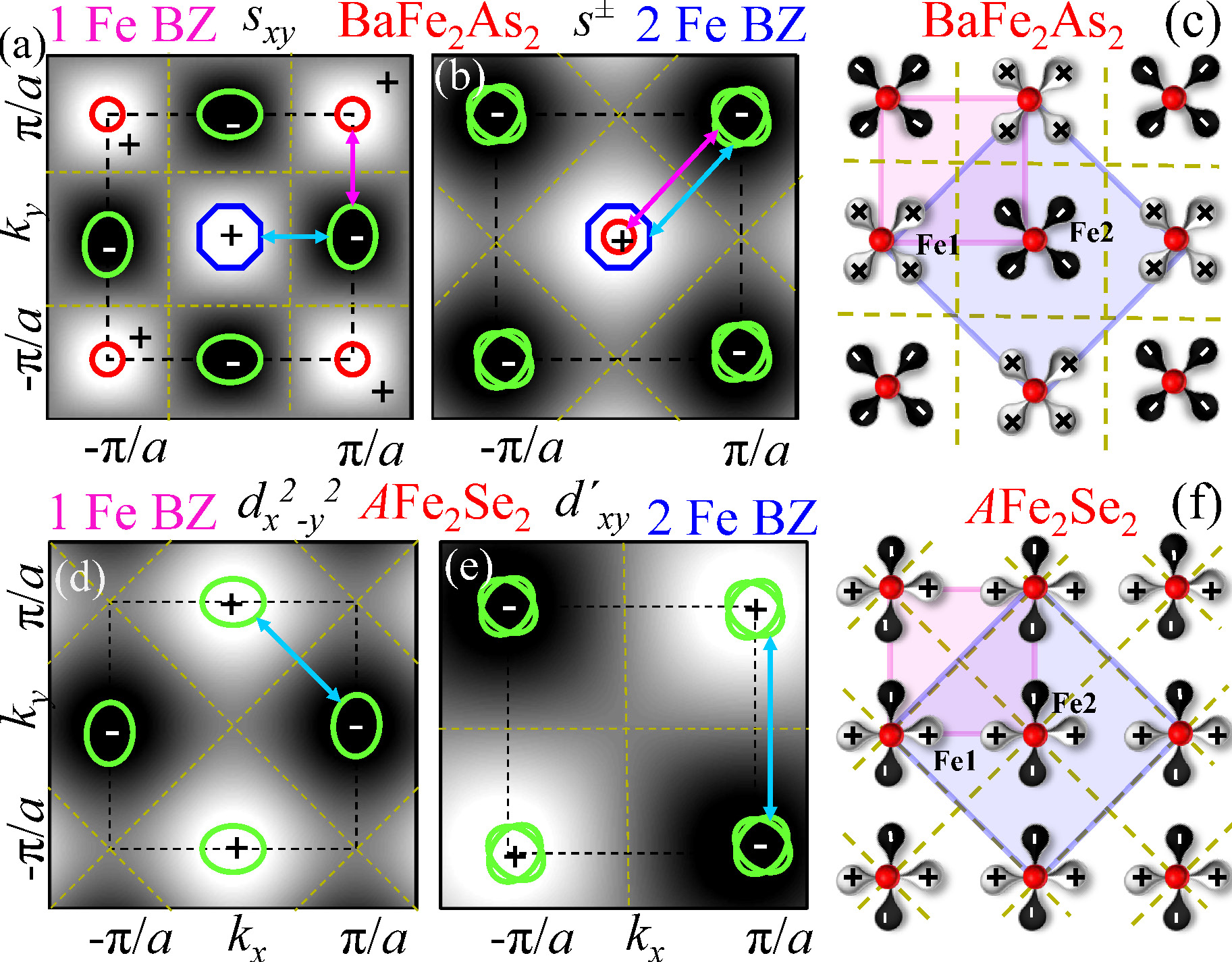}}
\caption{(Color online) Schematic behavior of the FS and pairing symmetry in 1 Fe per unit cell and 2 Fe per unit cell Brillouin zone for BaFe$_2$As$_2$ are shown in (a) and (b), respectively. The corresponding real-space view of the
 gap symmetry is depicted in (c) in which the SC phase is isotropic and nodeless on each Fe atoms but changes phase between Fe1 and Fe2 sublattices. Black to white color scale depicts the pairing symmetry from negative to positive in momentum space while black and white dumbbells represent positive and negative phase of the pairing symmetry in real space (exponential form of cosine and sine functions). Similar results for $A$Fe$_2$Se$_2$ compounds are given in (d)-(f). Although in momentum space FSs are isotropic and nodeless, but in real space the gap nodes pass through both of the Fe atoms. The phases of the pairing symmetries are $s_{xy}=2\cos{(k_xa)}\cos{(k_ya)}$, $s^{\pm}/d_{x^2-y^2}=\cos{(k_xa)}\pm\cos{(k_ya)}$, and $d_{xy}^{\prime}=2\sin{(k_xa)}\sin{(k_ya)}$.} \label{fig1}
\end{figure*}

We define a quantity in the 2 Fe unit cell by tilde. The unitary transformation consists of 45$^o$ rotation of the crystal with lattice constant $\tilde{a}\rightarrow\sqrt{2}a$ which gives $\tilde{k}_{x/y}=(k_x\pm k_y)/2$.\cite{footnote} In doing so, the $s^{\pm}$-pairing in the 2 Fe unit cell becomes a $\tilde{s}_{xy}$-like symmetry in the 1 Fe unit cell in which two inequivalent Fe sites posses opposite phases in real space, see Fig.~3(a-c). In momentum space, the SC gap symmetry is compatible with the FS topology. In real space, the gap phase obeys proper translational symmetry in the true 2 Fe unit cell, but not in the 1 Fe unit cell (conventional). This seems to pose a problem that is in fact an artificial issue. Proper unit cell contains 2 inequivalent  Fe sites; 1 Fe unit cell is an artificial one constructed for convenience of calculations assuming 2 inequivalent Fe to be the same.

We now look at the SC gap transformation. As obvious in pnictide, the sign change of the SC gap in $s^{\pm}$-pairing occurs between Fe1 and Fe2 atoms, see Fig.~3(c). Therefore, when the center Fe2 atom is removed from the lattice (when vacancies are present and ordered) in $A_x$Fe$_{2-x/2}$Se$_2$, %at $x\approx0.8$,
one can expect a different SC pairing symmetry to obtain sign change in the bulk SC order parameter. All of the Fe1 atoms are now equivalent, the sign change of gap in the pairing function should be carried by all atoms, i.e. a $d-$wave or $p-$wave state can arise. In$A_x$Fe$_{2-x/2}$Se$_2$ compounds when $T_c$ is highest at $x\sim0.8$, a $d-$wave gap is a most natural pairing state to arise.\cite{Das_FeSe} NMR studies also indicate the presence of spin-singlet pairing channel in the K intercalated samples with different composition.\cite{WYu_NMR} The $d-$wave pairing which gives rise to nodeless and nearly isotropic SC gap on the FSs, is also consistent with numerous observables.\cite{Das_FeSe}

On the other hand the $d-$wave SC gap in $A_x$Fe$_{2-x/2}$Se$_2$ compounds has an additional  complication. While a $d_{x^2-y^2}$-wave gap is compatible in both momentum and real space of the conventional 1 Fe unit cell, it is inconsistent when the pairing symmetry is transformed to the 2 Fe unit cell. For example, in Fig.~3(e), when a $k=(-\pi,-\pi)$ point is shifted by the reciprocal vector of $(2\pi/a,0)$ (assuming a square lattice), the FS topology is restored but the SC order maps on the site with  an opposite sign of the gap function. We propose that this problem can be cured by doubling the Brillouin zone in the momentum space. One way of doing it is by removing the Fe2 atom from the center of unit cell in Fig.~3(f). Therefore, a $d-$wave gap is possible in these compounds when the normal state vacancy order helps restore the translational symmetry of the crystal.

Iron-selenide materials host a SC dome-like feature centering at $x\approx 0.8$ as long as vacancy order persists, see Fig.~1(a). It has been reported that at $x=0.8$ all Fe2 atoms from the sublattice are completely removed,\cite{WBao_Tv}. This observation is consistent with the symmetry expected for the $d-$wave pairing. %Simultaneously, large saturated magnetic moments of 3.3$\mu_B$/Fe in a block checkerboard AFM order is also realized.\cite{WBao_Tv}
Such a pattern of competing order is forbidden in the 2 Fe 122-pnictide $I4/mmm$ structure\cite{WBao_common} and the system stabilizes to a low-temperature tetragonal $I4/m$ one, as indicated by the yellow box in Fig.~1(b).

In conclusion, we present a microscopic model to show that the onset of vacancy order leads to many significant modifications of electronic, magnetic and superconducting properties of iron-selenide superconductors. A block checkerboard AFM state with moments as large as 3.3$\mu_B$ is realized experimentally and is believed to be induced by the vacancy order at ${\bf Q}_m=4{\bf Q}_v$. Such a modulated ground state compensates for the broken translational symmetry of the $d-$wave superconductivity\cite{Das_FeSe} and therefore allows for the state with combined superconducting and structural modulations to develop. In this state superconducting gap is nodeless and nearly isotropic on the FSs, as observed in numerous experiments.

The results follow the same route as in other  high-$T_c$ superconductors where a modulated ground state of multiple competing phases promotes and coexists with superconductivity in a uniform phase rather than undergoing any nanoscale phase separation. The modulated superconductivity has also been seen in cuprates,\cite{Vojta,Sachdev} pnictides,\cite{Sasha_pnictide} URu$_2$Si$_2$\cite{Varma_URuSi,Sasha_URuSi} and Sr$_3$Ru$_2$O$_7$\cite{Borzi_SrRuO}, although the nature of competing phases may vary across these materials.

\begin{acknowledgments}
We thank W. Bao, R.S. Markiewicz, Z. Tesanovich, D. H. Lee, and S. Sachdev for useful discussions. This work is funded by US DOE, BES and LDRD and benefited from the NERSC computing allocations.
\end{acknowledgments}


\begin{thebibliography}{99}
\bibitem{foot3} We approximate all components of the potentials to be same for both bands, but neglecting interband components.
\bibitem{footnote} Such a unitary transformation is exact if two inequivalent Fe atoms become in-plane, otherwise it is an approximation. Such an approximation is reasonable as long the FS and gap symmetry are quasi-two-dimensional as in the present case.
\end{thebibliography}
\end{document}